\documentclass[jou]{apa6}

\title{Food Recommender Systems \\ \Large Important Contributions, Challenges and Future Research Directions}
\shorttitle{Food Recommender Systems}

\twoauthors{Christoph Trattner}{David Elsweiler}
\twoaffiliations{MODUL University Vienna \\ \url{christoph.trattner@modul.ac.at}}{University of Regensburg\\\url{David.Elsweiler@sprachlit.uni-regensburg.de}}

\abstract{The recommendation of food items is important for many reasons. Attaining cooking inspiration via digital sources is becoming evermore popular; as are systems, which recommend other types of food, such as meals in restaurants or products in supermarkets. Researchers have been studying these kinds of systems for many years, suggesting not only that can they be a means to help people find food they might want to eat, but also help them nourish themselves more healthily. This paper provides  a summary of the state-of-the-art of so-called food recommender systems, highlighting both seminal and most recent approaches to the problem, as well as important specializations, such as food recommendation systems for groups of users or systems which promote healthy eating. We moreover discuss the diverse challenges involved in designing recsys for food, summarise the lessons learned from past research and outline what we believe to be important future directions and open questions for the field. In providing these contributions we hope to provide a useful resource for researchers and practitioners alike.

}

\rightheader{Food Recommender Systems}
\leftheader{C. Trattner \& D. Elsweiler}

\usepackage[show]{chato-notes}
\usepackage{graphicx}
\usepackage{amssymb}
\usepackage{algorithm}
\usepackage{algorithmic}
\usepackage[table]{xcolor}
\usepackage{amsmath}
\usepackage[british]{babel}
\usepackage[natbibapa]{apacite}
\usepackage{url}
\usepackage{caption}

\newcommand{\specialcell}[2][c]{%
  \begin{tabular}[#1]{@{}l@{}}#2\end{tabular}}
  
  \usepackage{booktabs}

\usepackage{rotating}

\begin{document}
\maketitle    
                        
\section{Introduction}\label{sec:intro}

Online recommender systems have proved to be useful in diverse situations by
empowering the user to overcome the information overload problem, assisting with the decision making process and
serving as a means to change user behavior \citep{ricci2011introduction}. One domain, which has historically received comparatively little attention, however, especially when compared to areas relating to leisure and entertainment, is the recommendation of food items. This is surprising given the importance of food for human sustenance, the range of options available, the fact that making food choices is particularly challenging \citep{scheibehenne2010can}, and the high personal and societal costs of poor choices. Worldwide, lifestyle- and diet-related illnesses, such as obesity and diabetes, account for ~60\% of total deaths \citep{who2016a}. Both are conditions, which can be prevented and sometimes even reversed by appropriate dietary choices \citep{Ornish1990}.  

As such, health-aware food recommender systems are often mooted as an important
part of the solution to encourage healthier nutritional choices \citep{Freyne:2010:IFP:1719970.1720021,freyne2011recipe,harvey2012learning,Harvey:2013:YYE:2651320.2651339}.

There are many reasons, however, which make food recommendation challenging, not only in terms of encouraging healthy behaviour, but also in predicting what people would like to eat because this is complex, multi-faceted, culturally determined, not to mention context-dependent. Moreover, when developing food recommendation systems, there are additional issues for practitioners and researchers to consider, which do not arise in other recommendation domains. These include that users may have complex, constrained needs, such as allergies or life-style preferences, such as the desire to  eat only vegan or vegetarian food. In such cases, standard approaches work poorly and adequate data sources to filter recipes are not freely available. Other challenges include food items may have multiple names, ingredients can be prepared in different ways and unlike domains where products or media are recommended, it is not always clear if a recommended item can be prepared or consumed due to the potential for poor availability of ingredients, cooking  knowledge or equipment.

This paper makes two primary contributions. Firstly, we  
provide a summary of the state-of-the-art in food recommender systems, highlighting both seminal and most recent approaches to the problem, as well as important specializations, such as food recommendation systems for groups of users or systems which promote healthy eating. We examine which algorithms have been used in the food domain, how systems are typically evaluated, and the resources available to those interested in building or studying recommender systems in practice. In a second contribution we discuss the diverse challenges involved, as well as a summary of the lessons learned from past research and an outline of important future directions and open questions. In providing these contributions we hope to provide a useful resource for researchers and practitioners alike.

\begin{sidewaystable*}
\centering
\caption{Overview of different types of recommender systems strategies developed for recommending food (recipes, meal plans, groceries and menus) to people sorted in  chronological order by publication date.}
 \label{tbl:overview} 
\scalebox{0.65}{
\begin{tabular}{@{\extracolsep{0pt}}lllllllll} 
 \hline \\  [-2ex]
 \textbf{Author(s)} & \textbf{Algorithm(s)} &  \specialcell{\textbf{Person-}\\\textbf{alized}} & \specialcell{\textbf{RecSys}\\\textbf{Type(s)}} & \textbf{Feedback} & \specialcell{\textbf{Context/} \\ \textbf{Content} \\ \textbf{Feature(s)}} & \specialcell{\textbf{Dietary} \\  \textbf{Constrains}} & \textbf{Target} & \textbf{Dataset} \\   [2ex]\hline
 
  \citep{Elsweiler:2017:EFC:3077136.3080826} & \specialcell{Logistic \\ Random Forrest \\ Naive Bayes} & no & \specialcell{Recipes} & \specialcell{Ratings \\ Binary} & \specialcell{Title \\ Image \\ Ingredients \\ Nutrition \\ Pop. \& Appr} &  no&Single User & Allrecipes\\ 
  
   \citep{trattner2017investigating} & 
 \specialcell{LDA \\ WRMF \\ AR \\ SLIM \\ BPR \\ MostPop \\ User- ItemKNN} & yes/no & \specialcell{Recipes \\ Meal Plans} & \specialcell{Bookmarks \\ Ratings \\ Comments} & \specialcell{WHO-FSA \\ health score} & no&  Single User & Allrecipes\\ 
  
  \citep{Cheng:2017:ICS:3079628.3079641} & \specialcell{BPR \\ MostPop} & yes/no & \specialcell{Recipes} & \specialcell{Ratings} & \specialcell{City Size} &  no&Single User & Kochbar\\ 
  
   \citep{yang2017yum} & \specialcell{Learning to Rank} & yes & \specialcell{Recipes} & \specialcell{Binary} &  Image Embeddings & yes &  Single User & \specialcell{Yummly} \\ 

 \citep{Rokicki:2016:PPG:2930238.2930248} & \specialcell{UserKNN \\ MostPop} & yes/no & \specialcell{Recipes} & \specialcell{Ratings} & \specialcell{Gender} &  no&Single User & Kochbar\\ 

 \citep{Ge:2015:UTL:2750511.2750528} & \specialcell{MF \\ CB} & yes & \specialcell{Recipes} & \specialcell{Ratings \\ Tags} &  Tags & no&Single User & \specialcell{Wellbeing \\ Diet Book } \\ 

\citep{Elsweiler:2015:TAM:2792838.2799665} & \specialcell{SVD-Hybrid}  & yes & \specialcell{Meal Plans \\ (Set of recipes)} & Ratings & \specialcell{Ingredients}  & yes &  Single User & \specialcell{Quizine}\\ 
  
  \citep{sano2015recommendation} & \specialcell{UserKNN \\ SVD \\ Hybrid \\ NL-PCA} & yes & \specialcell{Groceries} & \specialcell{Purchases} & Food Categories & no &  Single User & \specialcell{Grocery \\ store data}\\ 
  
     \citep{Trevisiol:2014:BAR:2631775.2631784} & \specialcell{UserKNN \\ CB}  & yes & \specialcell{Menus \\ (Set of dishes)} & \specialcell{Binary} & \specialcell{Text \\ Sentiment}  & no &  Single User & \specialcell{Yelp}\\ 
    
  \citep{elahi2014interactive}& \specialcell{MF}  & yes & \specialcell{Recipes} & \specialcell{Ratings\\Tags} & \specialcell{tags}  & no &  Group of Users & \specialcell{Wellbeing \\ Diet Book }\\ 
  
    \citep{Harvey:2013:YYE:2651320.2651339}& \specialcell{CB, CF \\ Logistic Reg. \\ SVD-Hybrid}  & yes & \specialcell{Recipes} & Ratings & \specialcell{Ingredients \\ etc.}  & no &  Single User & \specialcell{Quizine }\\ 
    
     \citep{teng2012recipe} & \specialcell{SVM} & no & \specialcell{Recipes} &  \specialcell{Ratings} & \specialcell{Ingredients \\ Nutrition \\ Cook effort  \\ Cook methods } & no&Single User & Allrecipes\\ 

    \citep{kuo2012intelligent} & \specialcell{Graph-based\\ CB}  & yes & \specialcell{Menus \\ (Set of recipes)} & \specialcell{Tags} & \specialcell{Ingredients}  & no &  Single User & \specialcell{Food}\\ 

   \citep{el2012food}& \specialcell{CB \\ KB}  & yes & \specialcell{Food items} & \specialcell{Query} & \specialcell{tags}  & no &  Single User & \specialcell{USDA}\\ 
   
    \citep{  freyne2011personalized}& \specialcell{CF}  & yes & \specialcell{Meal plans \\ (Set of recipes)} & \specialcell{Ratings} & \specialcell{-}  & no &  Single User & \specialcell{Wellbeing \\ Diet Book}\\ 

    \citep{Ueta:2011:RRS:2186633.2186642}& \specialcell{KB}  & yes & \specialcell{Recipes} & \specialcell{Query} & \specialcell{tags}  & no &  Single User & \specialcell{Cookpad}\\ 

    \citep{vanPinxteren:2011:DRS:1943403.1943422} & \specialcell{CB}  & yes & \specialcell{Recipes} &  \specialcell{Cooked \\ recipes} & \specialcell{Recipe content \\features}  & no &  Single User & \specialcell{Smulweb}\\ 
    
      \citep{Freyne:2010:IFP:1719970.1720021} & \specialcell{UserKNN \\ CB \\ Hybrid} & yes & \specialcell{Recipes} & Ratings &  Ingredients &no& Single User & \specialcell{Wellbeing \\ Diet Book } \\ 
  
  \citep{berkovsky2010group} & \specialcell{UserKNN \\ GroupKNN \\ Hybrid}  & yes & \specialcell{Recipes} & Ratings & \specialcell{-}  & no &  Group of Users & \specialcell{Wellbeing \\ Diet Book }\\ 

  \citep{aberg2006dealing} &  \specialcell{CF} & yes & \specialcell{Meal Plans \\ (Set of recipes)} & \specialcell{Ratings} & - & yes &  Single User & \specialcell{Unknown}\\ 

  \citep{khan2003building} &  \specialcell{CBR} & yes & \specialcell{Meal Plans} & \specialcell{Query} & Nutrition Content & yes &  Single User & \specialcell{Unknown}\\ 
  
  \citep{mankoff2002using} & \specialcell{CB} & yes & \specialcell{Groceries} & \specialcell{Purchases} & Food groups & no &  Single User & \specialcell{Grocery \\ store data}\\ 
  
    \citep{lawrence2001personalization} &  \specialcell{AR \\ CF \\ CB} & yes & \specialcell{Groceries} & \specialcell{Purchases} & Product class & no &  Single User & \specialcell{Grocery \\ store data}\\ 

    \citep{hinrichs1991roles} &  \specialcell{CBR} & yes & \specialcell{Meal Plans} & \specialcell{Query} & Content & yes &  \specialcell{Single User\\Group of Users} & \specialcell{Unknown}\\ 
  
  \citep{hammond1986chef} &  \specialcell{CBR} & yes & \specialcell{Single New Recipe} & \specialcell{Query} & - & yes &  Single User & \specialcell{Unknown}\\ \hline

\end{tabular}
 
}
\end{sidewaystable*}

\section{Developed Approaches}\label{sec:approaches}

Despite food recommendation being a comparatively understudied problem in the research community, a decent body of literature exists. Table \ref{tbl:overview} provides a list of important research contributions relating to food recommendation. We list 25 popular, highly cited, recent and relevant papers in chronological order, selected using our experience in the domain in combination with bibliographic tools, such as Google scholar\footnote{\url{http://scholar.google.com}}, to identify the most relevant for the targeted readership. 
Special care was taken to identify work relating to different types of food item. As such, the papers and articles cited relate to the recommendation of recipes, meal plans, groceries and menus. Although the problem of recommending restaurants to people, e.g. \citep{park2008restaurant}, is related, especially when the meals served there are taken into consideration, we focus here on research relating to systems directly recommending the food items themselves.

The columns in Table \ref{tbl:overview} relate to dimensions that we believe characterize the nature of different contributions in the area. \textit{Algorithm} defines the various algorithmic approaches that have been tested in the food domain ranging from content based approaches, to collaborative filtering, to machine learning classifiers, some of which involve personalization (\textit{Personalized}). \textit{Recommended Items} describes the food item involved; \textit{Feedback} describes the means by which the system is informed on user preferences and the suitability of any recommendation provided; \textit{Context} provides the context dimension(s) utilised if applicable; \textit{dietary constraint} informs on whether nutrition was considered; \textit{Target} details who the end user(s) of the system was(were); and finally \textit{Dataset} details the proprietary or open dataset utilized. The remainder of the paper uses Table \ref{tbl:overview} as a structural basis.

In this section, we explain the approaches that have been taken in the literature to implement food item recommenders. In the literature the most prominent form of food recommender system provides single item recommendations mostly in the form of recipes. 

We structure the section around the approaches employed, summarizing content-based, collaborative filtering and hybrid approaches. We continue to show how context information is important and how this has been utilized in practice. Next, we broaden our focus to particular scenarios, which have been addressed, firstly looking at group-recommendations before reviewing research on food recommenders for healthy nutrition. 

Reflecting the literature as a whole, the majority of section details work recommending recipes to end users. That being said, there are other kinds of food items, which have been studied, albeit to a lesser extent. This is reflected in Table \ref{tbl:overview}. As datasets are becoming more readily available (see Section `Implementation Resources'), we expect interest in other food items to increase. The work published to date has largely employed the same standard approaches as have been applied to recipes. For example, 
content-based, collaborative filtering and hybrid approaches have been applied to restaurant review data to recommend menus \citep{Trevisiol:2014:BAR:2631775.2631784} and online shopping data to recommend groceries \citep{lawrence2001personalization,mankoff2002using,sano2015recommendation}.

\begin{figure*}[t!]
    \centering
    \scalebox{1} {
    \includegraphics[width=\linewidth]{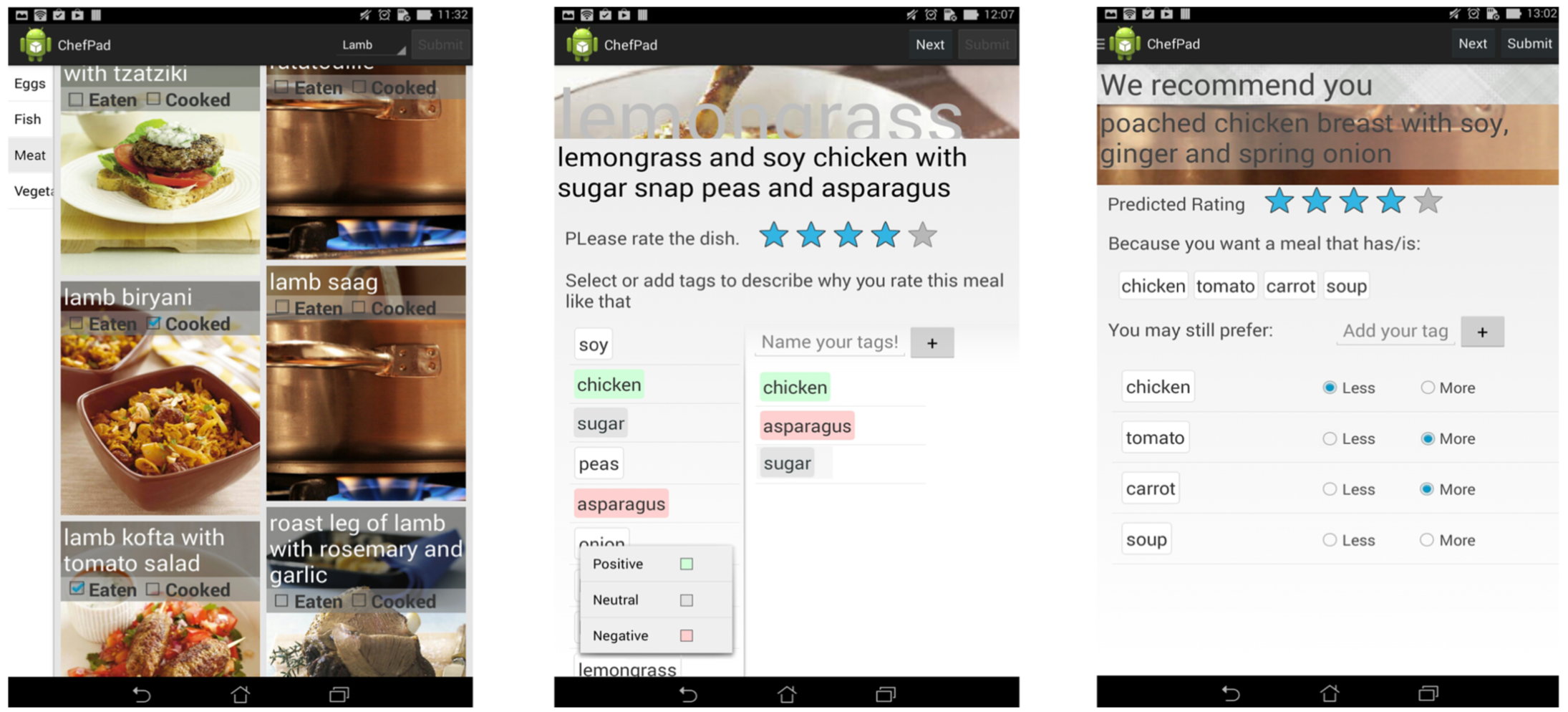}
    }
    \caption[]{ Example of a mobile food recommender interface as proposed by \protect\citet{elahi2014interactive} using not only ratings for preference elicitation but also tags at the same time. Taken with permission from the authors' work.}
    \label{fig:example_interface}
\end{figure*}

\subsection{Content-Based Methods (CB)}

Content-based approaches have been used as a means to tailor recommendations to the user's individual tastes. Freyne and Berkovsky, for example, made recommendations by breaking recipes down into individual ingredients and scoring based on the ingredients contained within recipes, which users had rated positively \citep{Freyne:2010:IFP:1719970.1720021,freyne2011recipe}. That is, if tomatoes had been present in recipes a user had reported liking, further recipes containing tomatoes would be predicted to also be liked by the user. Later work progressed this approach by not only accounting for positive ingredient biases, but also negatively weighting recipes based on contained ingredients featuring in recipes the user reported disliking \citep{Harvey:2013:YYE:2651320.2651339}.

\citet{teng2012recipe} proposed the use of complement and substitution networks as a means to generate  accurate predictions. Complement networks of ingredients are constructed via co-occurrence of the the same ingredients in the same recipes, while substitute networks are derived from user-generated suggestions for modifications. Experiments show that the use of these networks can predict the user preferences significantly better than approaches that rely on for example ingredient lists as features, cooking method, style, etc.

Other content-based approaches are more applicable to food recommender systems than other domains. For example, as food decisions are often visually driven \citep{mormann2012relative,schur2009activation}, the images associated with recipes can be exploited. Yang and colleagues have shown baseline approaches can be outperformed by algorithms designed to extrapolate important visual aspects of food images \citep{yang2015plateclick,yang2017yum}. In their work, Convolutional Neural Networks (CNN) provide a powerful framework for automatic feature learning.  \citet{Elsweiler:2017:EFC:3077136.3080826} also show that automatically extracted low-level image features, such as brightness, colorfulness and sharpness can be useful for predicting user food preference.

\subsection{Collaborative Filtering-Based Methods (CF)}

Collaborative filtering-based recommendation methods for food reccommender systems have also been proposed and evaluated. Freyne and Berkovsky tested a nearest neighbour approach using Pearson correlation on the ratings matrix, which offered poorer performance than the content approach described above \citep{Freyne:2010:IFP:1719970.1720021}. \citet{Harvey:2013:YYE:2651320.2651339} showed that SVD outperformed both the content and collaborative filtering approaches suggested in \citep{Freyne:2010:IFP:1719970.1720021}. \citet{Ge:2015:UTL:2750511.2750528} propose a matrix factorization (MF) approach for food recommender systems that fuses ratings information and user supplied tags to achieve significantly better prediction accuracy than content-based and standard matrix factorization baselines. They also present a mobile interface for the approach as shown in Figure \ref{fig:example_interface}. These screenshots show how a finer granularity of feedback can assigned via tags, complementing the standard binary and scaled ratings typically used.

More recently, 
 \citet{trattner2017investigating} tested a diverse range of collaborative filtering approaches implemented in the LibRec\footnote{\url{http://www.librec.net/}} framework using a large dataset crawled from the online recipe portal allrecipes.com. The highest performing CF approaches were Latent Dirichlet Allocation (LDA) \citep{griffiths2002gibbs} and Weighted matrix factorization (WRMF) \citep{hu2008collaborative}. The results of their experiments are shown in Table \ref{tab:rec_results2} below. 

\subsection{Hybrid Methods (Hybrid)}

Hybrid recommenders have been proposed by other scholars for the recipe recommendation task. For example,
\citet{Freyne:2010:IFP:1719970.1720021} combined a user-based collaborative filtering method with a content-base method. Moreover, in their follow-up work, which targeted groups of users (described in more detail in below), they employed a hybrid approach to combine three different recommender strategies in a single model, using a switching strategy. The switching was based on the ratio between the number of items rated by a user and overall number of items. Another example of a hybrid approach can be found in the work of \citet{Harvey:2013:YYE:2651320.2651339} who achieved the best performance in their experiments by combining an SVD approach with user and item biases.
 
\subsection{Context-Aware Approaches}\label{subsec:context}

Numerous exploratory data analyses have demonstrated that context is important in food recommendation, with gender \citep{Rokicki:2016:PPG:2930238.2930248}, time \citep{kusmierczyk2015temporal}, hobbies \citep{Trattner:2017:RCI:3099023.3099072}, location \citep{Cheng:2017:ICS:3079628.3079641,DeChoudhury:2016:CDC:2818048.2819956, zhu2013geography} and food availability \citep{DeChoudhury:2016:CDC:2818048.2819956} being identified as important variables. All of these studies employed relatively simple filtering techniques to split naturalistic datasets in order to explore how recipes were rated \citep{freyne2011personalized}, bookmarked \citep{trattner2017investigating}, 
shared \citep{abbar2015you} or relate to health statistics \citep{trattner2017monitoring}.

\citet{harvey2012learning} collected detailed context data encapsulating the ratings participants provided for their dataset, where participants could identify a broad range of factors to justify the rating assigned to a recipe as a meal to cook for dinner that day. Analyzing these with regression modeling showed that factors, such as how well the preparation steps are described, as well as the nutritional properties of the dish, the availability of ingredients and temporal factors such as day of the week have a bearing on the user's opinion of the recommendation.

What is lacking with respect to context is an understanding of which variables are the most important and how best to account for these algorithmically \citep{rokicki2017editorial}. Despite studying numerous factors, \citet{Harvey:2013:YYE:2651320.2651339}, for example, limited their algorithmic efforts to approaches with only nutritional, user and item biases.

In summary, although the problem of improving the precision of recommendations has been attended to by numerous researchers with diverse approaches, the results achieved for the recommendation of recipes to individual users, measured on standard metrics are typically poorer than in other domains. To demonstrate typical results achieved with standard approaches on recipe data, we present the results of our own experiments with standard techniques on a well-known dataset in Table \ref{tab:rec_results2} below. These results underline the challenge of predicting which dishes people will like and emphasize that further effort is required.

\subsection{Group-Based Methods}

Of course people do not always eat or make food choices alone. Often these are activities done together with friends, families or colleagues. It is well known in social psychology that the social situation within which one will eat (who is present, why they are present, and what their preferences are) influences the food choices taken \citep{wansink2006mindless}. In food recommender systems, such social contexts are addressed by group recommender systems. In this setting, a list of items is produced for a group of people rather than for an individual user. Despite the pervasiveness of shared food consumption experiences, group-based food recommender systems research has been limited, even though the earliest efforts can be traced to the early 1990's  \citep{hinrichs1991roles}.  \citet{berkovsky2010group} not only studied different strategies for recommending recipes to a group of people but evaluate these methods with real users in a family scenario. In particular their work introduces four different strategies:  A general strategy (which employs a most popular approach to recommend items), an aggregated model (which first combines individual user models into a single model before applying the collaborative filter), aggregated predictions strategies (which first computes CF on the individual user profiles and then combines the predicted rating) and finally a personalized strategy (which exploits a standard CF algorithm). The results show that the personlized version works the best but it was not possible to create personalized recommendations for all of the users. 
More recently \citet{elahi2014interactive} proposed a  mobile interface and algorithm for food recommender system in a group-context. In addition to improving the prediction algorithm with tags, the authors use group-based preference elicitation, in which users play different roles in the food choice process. One user is designated as the group leader or cook to whom the system delivers meal recommendations based on the group utility score, which aggregates predictions using the tags and ratings of all the group members.

\subsection{Health-Aware Methods}\label{subsec:health}


When motivating research on food recommender systems, health problems and improving nutritional habits are usually mentioned e.g. \citep{Freyne:2010:IFP:1719970.1720021,freyne2011recipe,harvey2012learning,Harvey:2013:YYE:2651320.2651339}. Incorporating health into the recommendation, however, has largely been a recent focus \citep{Schafer:2017:THR:3079452.3079499,DBLP:conf/recsys/ElsweilerHLSSTT17,DBLP:conf/recsys/ElsweilerLSST16}. 
One means of achieving this is to incorporate nutritional aspects into the recommendation approach directly.  \citet{Ge:2015:HFR:2792838.2796554} took this approach by accounting for calorie counts in the recommendation algorithm.
They did this based on a so-called ``calorie
balance function'' that accounts for the differences between the calories the user needs and the calories in a recipe.

\citet{elsweiler2015bringing} refer to the trade-off for most users between recommending the user what she wants and what is nutritionally appropriate. This is a trade-off applicable for a large proportion of users \citep{Harvey:2013:YYE:2651320.2651339} and should be optimized \citep{elsweiler2015bringing}. The authors proposed combining to two aspects linearly as a framework for evaluating different algorithmic approaches to incorporate health in the recommendation process. 

The formula (see Equation \ref{eq:1}) illustrates the simple concept. Here, $i$ is a given recipe, $\hat{r(i)}$ is the estimated rating for recipe $i$, Max($\hat{r(i)}$) is the maximum estimated rating over all recipes. $n(i)$ is the nutritional ``error'' incurred when recommending this recipe (relative to some ideal set of nutritional values). $\lambda$ is a free parameter that can be set to suit the researcher/practitioner's priorities, although $\lambda$=.5 is probably preferable initially as it gives equal weighting to rating and nutrition. Note that all of these estimates are implicitly conditioned on a specific user $u$.
\begin{equation}\label{eq:1}
score(i) = \lambda \frac{r(i)}{max(r(i))}+(1-\lambda)-1\times\frac{n(i)}{max(n(i))}
\end{equation}

\citet{trattner2017investigating} employed a post-filtering (see Equation \ref{eq:2} and \ref{eq:3}) approach to incorporate further nutritional aspects. To post-filter items a
a straightforward scoring function is applied which re-weights the scores of a recipe for a particular user based on the WHO or inverse FSA score, employing a simple multiplication. The $score_{u,i}$ in the equation stands for the score of the item $i$ for user $u$ and $who_i$, $fsa_i$  denote the health scores for that item. The two nutrition metrics are based on widely accepted nutritional standards from The World Health Organisation (WHO) \citep{who2003diet} and the United Kingdom Food Standards Agency (FSA) \citep{fsa2007} (see Section `Implementation Resources'). Their previous work had used these measures to establish the (un)healthiness of recipes from a popular Internet food portal \citep{trattner2017}.
\begin{equation}\label{eq:2}
score_{u,i,who} = score_{u,i} 	\cdot (who_i+1)
\end{equation}
\begin{equation}\label{eq:3}
score_{u,i,fsa} = score_{u,i} 	\cdot (16-fsa_i-4+1)
\end{equation}

\begin{table*}[t!]
\centering
\caption[]{Recommender ranking accuracy sorted by nDCG and recommender accuracy post-filtered by FSA scores. The mean FSA scores of the top-5 recommended recipes are also reported along with the different average nutriens of the lists and the according FSA health labels (taken from \protect\citet{trattner2017investigating}).}
 \label{tab:rec_results2}
\scalebox{1}{
\begin{tabular}{@{\extracolsep{0pt}}lcccccc} 
 \hline \\  [-2ex]
\textbf{Algorithm} & \textbf{nDCG@5} & \specialcell{\textbf{FSA} \\ \textbf{score}} & \textbf{Fat (g)} & \textbf{Sat. Fat (g)} & \textbf{Sugar (g)} & \textbf{Sodium (g)}\\  \hline
LDA & .0395 & 9.110  & 8.70	\cellcolor{orange!25} & 3.73	\cellcolor{orange!25} & 8.73	\cellcolor{red!25} & 0.32	\cellcolor{orange!25}\\ 
WRMF & .0365  & 9.114 & 9.50	\cellcolor{orange!25} & 3.89	\cellcolor{orange!25} & 8.84	\cellcolor{red!25} & 0.34	\cellcolor{orange!25} \\ 
AR  & .0343  & 9.206  & 9.27	\cellcolor{red!25} & 4.12	\cellcolor{orange!25} & 10.50	\cellcolor{orange!25} & 0.25	\cellcolor{orange!25} \\ 
SLIM & .0326  & 8.907  & 9.27	\cellcolor{orange!25} & 3.82	\cellcolor{orange!25} & 7.91	\cellcolor{red!25} & 0.33	\cellcolor{orange!25} \\ 
BPR & .0325  & 9.252  & 8.69	\cellcolor{orange!25} & 3.82	\cellcolor{orange!25} & 7.83	\cellcolor{red!25} & 0.29	\cellcolor{orange!25} \\ 
MostPop & .0294  & 9.004 & 9.02	\cellcolor{orange!25} & 3.94	\cellcolor{orange!25} & 10.01	\cellcolor{orange!25} & 0.23	\cellcolor{orange!25} \\ 
UserKNN & .024  & 8.985 & 8.96	\cellcolor{orange!25} & 3.73	\cellcolor{orange!25} & 7.98	\cellcolor{red!25} & 0.31	\cellcolor{orange!25} \\ 
ItemKNN  & .0178  & 8.652  & 8.59	\cellcolor{orange!25} & 3.51	\cellcolor{orange!25} & 6.03	\cellcolor{red!25} & 0.31	\cellcolor{orange!25} \\ 
Random  & .0029  & 8.486  & 8.74	\cellcolor{orange!25} & 3.49	\cellcolor{orange!25} & 5.71	\cellcolor{red!25} & 0.30	\cellcolor{orange!25} \\ \hline 
\\ [-1.5ex]
  & \multicolumn{6}{c}{FSA score post-filtered ($score_{u,i,fsa}$)}   \\  [0.5ex] \hline
LDA  & .0321  & 7.323  & 6.51	\cellcolor{green!25} & 2.42	\cellcolor{orange!25} & 4.03	\cellcolor{orange!25} & 0.29	\cellcolor{orange!25} \\ 
WRMF & .0303  & 7.361  & 6.48	\cellcolor{green!25} & 2.30	\cellcolor{orange!25} & 4.75	\cellcolor{orange!25} & 0.31	\cellcolor{orange!25}\\ 
SLIM & .0248  & 7.008  & 6.20 \cellcolor{green!25} & 2.56	\cellcolor{orange!25} & 2.59	\cellcolor{orange!25} & 0.24	\cellcolor{orange!25} \\ 
AR & .0238 & 6.984  & 5.64	\cellcolor{green!25} & 1.94	\cellcolor{orange!25} & 3.95	\cellcolor{orange!25} & 0.28	\cellcolor{orange!25} \\ 
MostPop & .0228  & 7.334  & 5.37 \cellcolor{green!25} & 2.02	\cellcolor{orange!25} & 2.46	\cellcolor{orange!25} & 0.24	\cellcolor{orange!25} \\
BPR  & .0205  & 6.722  & 6.42	\cellcolor{green!25} & 2.30	\cellcolor{orange!25} & 4.95	\cellcolor{orange!25} & 0.26	\cellcolor{orange!25} \\ 
UserKNN  & .0168  & 6.722  & 6.88	\cellcolor{green!25} & 2.73	\cellcolor{orange!25} & 3.33	\cellcolor{orange!25} & 0.33	\cellcolor{orange!25} \\ 
ItemKNN  & .0109  & 6.124  & 5.15	\cellcolor{green!25} & 1.79	\cellcolor{orange!25} & 3.51	\cellcolor{green!25} & 0.25	\cellcolor{orange!25} \\ 
Random  & .0022  & 4.305  & 1.59	\cellcolor{green!25} & 0.43	\cellcolor{green!25} & 1.45	\cellcolor{green!25} & 0.09	\cellcolor{green!25} \\ \hline

\end{tabular}
}

\end{table*}

Table \ref{tab:rec_results2} describes the performance of 9 prominent recommender algorithms as implemented in the LibRec framework in Trattner and Elsweiler's experiments.  The top and bottom halves of table shows the performance without and with post-filtering respectively. Full details of the experimental setup can be found in \citep{trattner2017investigating}.

These experiments with post-filtering on nutritional properties show that 1) it is possible to balance and potentially optimize the trade-off between recommendation accuracy and the healthiness of recommendations, 2) some recommendation algorithms may be more (e.g., LDA and WRMF) or less suitable (e.g., MostPop and BPR) to this process.

Nevertheless the results also show that 3) while the approach shows potential benefit and future work should try to optimize the trade-off, the method by itself will not lead to healthy nutrition - at least not with the collection evaluated in this work. Despite offering a significant improvement on the standard approaches, the post-filtered results show that the best FSA and WHO scores achieved were not particularly high and are associated with extremely poor recommendation accuracy. These represent the best health values which can be achieved using an individual item recommendation approach, indicating that complementary ideas are necessary.

One such complementary approach is to combine individual recommended items for a user, such that they meet the recommended intake for that user over a longer period of time (e.g. day, week etc.). \citet{freyne2011personalized} presented an interface, which allowed users to generate their own meal plans from individually recommended dishes. The recommendations were generated using the authors' hybrid approach as described above \citep{Freyne:2010:IFP:1719970.1720021}. The interface for such plans evaluated on 5000 people in Australia. To encourage variation in meal plans a decay function was applied to meals appearing regularly in plans. Users manually created plans from lists of recommendations but the lists were filtered such that only meals that could be added and still ensure plans met guidelines featured in the list of recommended items.

\citet{harvey2015automated} presented a similar interface, which automated the creation of plans consisting of a combination of breakfast, lunch and dinner plus an allowance for snacks and drinks. The same authors evaluated their planning approach systematically by deriving plans from taste profiles (i.e. from users featuring in naturalistic datasets) combined with diverse personas (simulated user properties, such as height, weight, gender, age, nutritional goal (lose/gain/maintain weight) and activity level (from sedentary to highly active) \citep{Elsweiler:2015:TAM:2792838.2799665}. In a first step, the authors estimated ratings users with particular profiles might assign to recipes (using approaches like those described above). In a second step, following approaches from nutritional science, the recommended nutritional intake was calculated for the user persona, including the required calories, but also where these should be sourced (proteins, carbohydrates etc.). Lastly, plans were generated for a given user (persona-profile combination) by taking the top-n recommendations from the recommender system for the taste profile, splitting these into two separate sets, one for breakfasts and one for main meals and performing a full search finds \textit{every combination} of these recipes in the sequence [breakfast, main meal, main meal] meeting the target nutritional requirements as defined above. 

Using this method the authors were able to generate plans for 4025/6400 cases (63\%) and at least 1 plan was generated for 58 out of the 64 (91\%) user profiles and for each of the 100 personas. The authors moreover analyzed the factors, which made the development of plans challenging. When personas required a relatively high calorie intake, e.g. if the persona was tall or wanted to gain weight, the simple approach using 3 meals of fixed portions was often unable to address this properly. Similarly, profiles with little diversity in preferred ingredients were also hard to satisfy.

Substituting meals has been mooted as a further approach to influencing food choices. \citet{Elsweiler:2017:EFC:3077136.3080826} developed predictive models with the aim of forecasting the choices people will make. After evaluating the models for prediction accuracy using cross-validation,  these were used to select recipe replacements such that users were be ``nudged'' towards making healthier choices. Aligning with the findings reported above, visual and nutritional features were important.  A user study found that using the predictive models as the basis for recommendations, participants were significantly more likely to choose a recipe with much less fat content - the opposite of the trend that one typically sees.

Substituting ingredients within recipes has also been proposed to improve the health credentials of individual recipes healthier e.g.  \citep{teng2012recipe,AchananuparpW16}. This approach has, however, yet to be evaluated properly in a nutritional context. Initial steps  in this direction were taken by \citet{KusmierczykHT}, whose findings illustrate to what extent it may be possible to recommend a user substitute ingredients based on the user's previous recipe uploads and  accounting for social-, temporal and geographic-context.

In our experience the standard approaches applied to date in the literature do not work well when dealing with specialist diets, e.g. (vegetarian, vegan or allergies)\footnote{We have not published our findings, but we have run several test runs with vegetarian, vegan, and gluten allergy user profiles.}. Constraint-based approaches are found surprisingly rarely in the literature.

One exception is \citet{yang2017yum} who had access to the data basis to apply filters based on vegetarian, vegan and gluten-free food. A further example can be found in the nutritional science literature whereby linear programming is used to ensure Malawian children achieve the required nutritional intake recommended by experts \citep{ferguson2004food}. As comparable datasets exist (see Section `Implementation Resources' below), there is no reason why a similar approach cannot be taken to promote healthy eating patterns in other demographics.

\section{Addressed Challenges and Problems}\label{sec:challenges}

As should now be clear the food recommendation task brings additional challenges to those in other recsys domains. There are also standard challenges, applicable to all domains, which have been addressed, at least to some extent, in food recommender research. In this section we first relate the generic challenges and how these have been addressed or not in the food domain, before switching focus to the challenges unique to food recommendation.

\textit{User preference sources.} Food recommendation research has mainly exploited explicit sources of user feedback in the form of ratings \citep{freyne2011personalized,Freyne:2010:IFP:1719970.1720021,Harvey:2013:YYE:2651320.2651339}, bookmarks \citep{trattner2017investigating} or shares \citep{abbar2015you}. Methods of implicit feedback have been used less often, but examples include recipe views \citep{west2013cookies,wagner2014nature} and the sentiment of reviews submitted about recipes \citep{trattner2017investigating}. 

\textit{User preference scarcity.} To our knowledge the problems of scarcity of user feedback, illustrated by the cold-start problem and sparse matrices, has not been directly addressed in the food recommender systems literature. Rather standard  solutions, which cope well, such as SVD have been applied \citep{Harvey:2013:YYE:2651320.2651339,trattner2017investigating}.   

\textit{Offline and online evaluation of recommendations.} To our knowledge, evaluation in the food recommendation domain has been almost offline. Typically, as is explained in more detail below, datasets have been created naturalistically e.g. \citep{yang2017yum,Trevisiol:2014:BAR:2631775.2631784,trattner2017investigating} or via user studies \citep{Freyne:2010:IFP:1719970.1720021,Harvey:2013:YYE:2651320.2651339}. These datasets form the basis of offline evaluations in the form of prediction tasks. Other evaluations have taken the form of user studies, where users test interfaces in a semi-controlled \citep{Ge:2015:UTL:2750511.2750528} or naturalistic environment \citep{freyne2011personalized}. However, full-online evaluations have to our knowledge not yet been published. 

\textit{Beyond accuracy.} Accuracy has been the overwhelming focus of research efforts to date but nevertheless, as described above, it remains a challenge, which in the food domain, has yet to be adequately solved. Accuracy, however, is not the only important aspect to consider when recommending food. Novelty and serendipity are both properties of food recommendations, which users appreciate \citep{Harvey:2013:YYE:2651320.2651339}, but to our knowledge, these are yet to be studied. \citet{Elsweiler:2015:TAM:2792838.2799665} did acknowledge the importance of dietary diversity in their meal plan work. Moreover, the preference-healthfulness trade-off bears many similarities to traditional work on novelty and serendipity in that it involves recommending non-preferred items while minimizing the loss in precision. While preliminary research in this direction exists \citep{Elsweiler:2015:TAM:2792838.2799665,trattner2017investigating}, there is much work to do in order to understand how to optimize this trade-off appropriately. 

\textit{Recommendation visualizations and explanations.} Methods of visualization and the explanation of recommendations have been, at best, implemented in a superficial way within food recommender research. Examples include the traffic light system employed by \citet{trattner2017investigating} and the plan meta-data provided in the demo system presented by \citet{harvey2015automated}. \citet{elahi2014interactive} provide the best example of explanations for the recommendations offered by their system as can be seen in  in Figure \ref{fig:example_interface}. Nevertheless, only superficial evaluations of any of these systems have been published.

\textit{Other common challenges.} Despite their importance generally to recommender systems, there is nothing to report from the food domain in terms of significant contributions on the issues of  privacy and collaborative recommenders, scalability and distribution of collaborative recommenders or issues of robustness or attacks on food recommenders.			

\textit{Challenges unique to food recommender systems.} 	
We can see from the numerous challenges yet to be addressed in the food domain, that research in this area is still preliminary. That being said, we wish to acknowledge some domain specific challenges, which have been addressed to some extent. Firstly, as Section `Developed Approaches' shows, the challenge of tailoring standard approaches to the problem has been tackled. 

There have been efforts to better process and understand the content of items to be recommended. These include normalizing ingredients and ingredient quantities \citep{muller2012estimating,kusmierczyk2015temporal};
understanding the role of context in user decision processes (see Section on context-aware recommender systems), and understanding which visual features are helpful in guiding these choices \citep{yang2017yum,Elsweiler:2017:EFC:3077136.3080826}.

With respect to health, there have been preliminary efforts to model nutritional aspects of the process \citep{schafer2017user}, which include user requirements \citep{gibney2002introduction}, user intake \citep{strassburg2010ernahrungserhebungen} and the estimation of portion sizes \citep{zhang2011food}. Other work has pre-processed recipes to establish the nutritional content either by ingredient matching \citep{muller2012estimating} or by visually analyzing food images \citep{chokr2017calories}. Finally, as we described in detail above, progress has been made in incorporating health in the recommendation process either by considering nutrition in item recommendation e.g.  \citep{Ge:2015:UTL:2750511.2750528}, generating meal plans \citep{Elsweiler:2015:TAM:2792838.2799665} or via algorithmic nudging \citep{Elsweiler:2017:EFC:3077136.3080826}. It is unclear, however, which method works most effectively.

\section{Implementation Resources}\label{sec:impl_resources}
In this section we summarize resources that can help in the development of food recommender systems. We summarize (i) datasets typically used to study food consumption patterns and to evaluate algorithmic approaches, (ii) nutrition and health resources, available to implement health-aware recommender systems. Finally, frameworks typically employed to build these are described.

\subsection{Recipe, Meal plan, Menu and Grocery Store Datasets} 
To date research in the food recommender systems domain typically relies on proprietary and none standardized datasets. This contrasts with domains such as movie recommender domain, where the well-known MovieLens datasets have set a standard. The following list highlights datasets usually employed when it comes to the implementation of recipe, meal plan, grocery and menu recommender systems.

\textit{Recipes.} Most of the research for recommending recipes relies on Web resources, e.g., Allrecipes\footnote{\url{http://www.allrecipes.com}} or Food.com\footnote{\url{http://www.food.com}} which comprise rich item and user profiles. Although these offer an extensive basis for conducting research in that direction, most of the datasets cannot be shared as the terms of services of the sites explicitly forbid it. As such, few publicly accessible datasets comprising recipe and user profiles are available.  Researchers must typically develop their own crawlers or seek a license agreement with the platform providers. The Australian government agency CSIRO'S Wellbeing Diet Book\footnote{\url{https://www.csiro.au/en/Research/Health/CSIRO-diets/CSIRO-Total-Wellbeing-Diet}} has been used by Australian researchers \citep{Freyne:2010:IFP:1719970.1720021} 
and connected researchers in Italy \citep{elahi2014interactive}, but is not readily available to other researchers.  Cookpad\footnote{\url{http://www.cookpad.com}} and Yummly\footnote{\url{http://www.yummly.com}} have both supported academic research by providing licensed access to recipe and profile data, and Yummly also supports broad access to restricted  data via a no-cost API. One dataset has recently been made available by the Massachusetts Institute of Technology (MIT)\footnote{\url{http://im2recipe.csail.mit.edu}} comprising of over 1 million recipes including food images and some meta-data. The dataset is limited, however, in that no user profiles or interactions are available, and as such the dataset may not be suitable for evaluating a recipe recommender system in an offline scenario.  
The lack of standard collections restricts the reliability and generalizability of research published to date.

\textit{Meal plans and restaurant menus.} Meal plan recommender research has typically relied on the same recipe datasets as above.  To our knowledge no freely available datasets containing meal plans exist.  Yelp\footnote{\url{http://www.yelp.com}} has been used as a resources to build and evaluate menu recommender system algorithms. As with recipe datasets, in order to obtain the data one might need to implement a crawling framework as the terms of services of the site to date omit data sharing

\textit{Groceries.} In the grocery recommender scenario, to our knowledge, only one dataset is freely available. This dataset was published by Kaggle\footnote{\url{https://www.kaggle.com/c/instacart-market-basket-analysis}} and contains 3 millions purchases of users on instacart and comprises limited meta-data (such as grocery name) in respect to the groceries bought out in a basket. Table \ref{tbl:overview} refers to some other datasets but these are not available publicly.
 
\begin{table}[t!]
\centering
\caption{Example of an nutrition entry for the query `apple' in the USDA database.}
 \label{tab:example}
\scalebox{1}{
\begin{tabular}{@{\extracolsep{0pt}}lll} 
 \hline \\  [-2ex]
\textbf{Nutrition} & \textbf{Unit} & \textbf{Value per 100g} \\ [0.5ex] \hline
Water & g & 85.56 \\
Energy & kcal & 52 \\ 
Protein & g & 0.26 \\
Total lipid (fat) & g & 0.17 \\
Carbohydrate, by difference & g & 13.81 \\
Fiber, total dietary &  g & 2.4\\
Sugars, total &  g & 10.39 \\ \hline
\end{tabular}
} 
\end{table}

\subsection{Nutrition \& Health Resources} \label{subsec:nutr}

When it comes to the implementation of food recommender systems or algorithms it is not only beneficial to have open-data datasets comprising of user and item profiles (as discussed earlier), but also other external resources that help in building such a system. For instance, to build a health-aware recipe recommender system, it is essential to know the nutritional values of food items and to what extent these may be healthy or unhealthy
To estimate nutrition, the typical approach is to map ingredients to standard databases, such as those provided by the USDA\footnote{\url{https://ndb.nal.usda.gov/ndb}} (US) or the BLS\footnote{\url{https://www.blsdb.de}} (Germany). As an example, Table \ref{tab:example} provides a partial entry for the ingredient  `apple'\footnote{\url{https://ndb.nal.usda.gov/ndb/foods/show/2122}}. The example is far from being complete, as also `Minerals', `Vitamins', `Lipids' and other macro nutrients can be obtained such as `Caffeine' are also accessable in the database. One of the challenges typically involved in the matching process is the normalization of the ingredients in a recipe, as different names are often used to express the same entity, such as `100g Parmesan cheese' vs `100g of shredded Parmesan cheese'. The method of processing or cooking may additionally influence the nutritional value. Moreover, units are often not expressed using normalized units of quantity. One recipe may refer to `one cup of water' whereas another may refer to the same item as `235ml water'. Detailed descriptions of the challenges involved can be found in \citep{muller2012estimating}. Standard NLP techniques such as stop-word removal, conjunction splitting, string matching, etc. can be applied to address some of these (see for example \citep{kusmierczyk2015temporal}). A more practical means to extract this kind of information is though to for instance employ a Web services such as provided by \textsc{Spoonacular}\footnote{\url{https://market.mashape.com/spoonacular}}, whose API is able to extract ingredient names and amounts in a unified way, which can in some cases be accessed for free for purposes of academic research.

\begin{table*}[t!]
\centering
\caption[]{FSA front of package guidelines as proposed in \protect\citet{fsa2007} and as, for example, used in \protect\citet{trattner2017investigating}.}
 \label{tab:fsa}
\scalebox{1}{
\begin{tabular}{@{\extracolsep{0pt}}llll} 
 \hline 
\textbf{Text} & \cellcolor{green!25} \textbf{LOW} & \cellcolor{orange!25} \textbf{MEDIUM} & \cellcolor{red!25} \textbf{HIGH}\\
\textbf{Color code} &  \cellcolor{green!25} \textbf{Green} & \cellcolor{orange!25} \textbf{Amber} & \cellcolor{red!25} \textbf{Red}\\ \hline
Fat &  \cellcolor{green!25} $\leq$ 3.0g/100g & \cellcolor{orange!25} $>$ 3.0g to $\leq$ 17.5g/100g & \cellcolor{red!25} $>$ 17.5g/100g or $>$ 21g/portion\\
Saturates & \cellcolor{green!25} $\leq$ 1.5g/100g & \cellcolor{orange!25} $>$ 1.5g to $\leq$ 5.0g/100g & \cellcolor{red!25} $>$ 5.0g/100g or $>$ 6.0g/portion\\
Sugars & \cellcolor{green!25} $\leq$ 5.0g/100g & \cellcolor{orange!25} $>$ 5.0g to $\leq$ 22.5g/100g & \cellcolor{red!25} $>$ 22.5g/100g  or $>$ 27g/portion\\
Salt & \cellcolor{green!25} $\leq$ 0.3g/100g & \cellcolor{orange!25} $>$ 0.3g to $\leq$ 1.5g/100g & \cellcolor{red!25} $>$ 1.5g/100g or $>$ 1.8g/portion\\ \hline
\end{tabular}
}
\end{table*}

\begin{table*}[t!]
\centering
\caption{WHO guidelines as originally proposed in \protect\citet{who2003diet} and adopted to recipes by \protect\citet{howard2012nutritional} and as, for example, used in \protect\citet{trattner2017investigating}.}
 \label{tab:who}
\scalebox{1}{
\begin{tabular}{@{\extracolsep{0pt}}lc} 
 \hline \\  [-1.5ex]
\textbf{Dietary Factor} & \textbf{Range (percentage of kcal per meal/recipe)} \\ [0.5ex] \hline
Protein & 10-15 \\
Carbohydrates & 55-75 \\
Sugar & $<$ 10 \\
Fat & 15-30 \\
Saturated Fat & $<$ 10 \\
Fiber density (g/MJ) & $>$ 3.0$^{\dagger}$ \\
Sodium density (g/MJ) & $<$ 0.2$^{\ddagger}$ \\ \hline
\multicolumn{2}{l}{$^{\dagger}$Based on 8.4 MJ/day (2,000 kcal/day) diet and recommended daily  fiber intake of $>$25g.}\\
\multicolumn{2}{l}{$^{\ddagger}$Based on 8.4 MJ/day (2,000 kcal/day) diet and recommended daily sodium intake of $<$2g.}
\end{tabular}
}
\end{table*}

Other resources to identify the nutritional properties of a meal (recipe) are provided by \citep{muller2012estimating}. These output the nutritional properties for a given German recipe by utilizing the BLS database. M\"uller employ a multi-step process, first utilising a rule-based infrastructure before a learning to rank approach to identify the most appropriate database entry for a given ingredient. The framework can be obtained from the authors without cost but a license for the BLS is required to use the software.  The \textsc{Edamam}\footnote{\url{https://www.edamam.com}} Web service offers similar functionality for English and Spanish recipes. This service is a commercial product, but as with Spoonacular, can in some cases used without cost for academic purposes.

To estimate the healthiness of a meal \citep{trattner2017}, one may rely on standards as set by nutrition scientists. There are many of such standards for different countries and other geographical regions. The ones which have been successfully applied to the food recommender problem (see \citep{trattner2017investigating,Elsweiler:2017:EFC:3077136.3080826}) are provided by the Food Standard Agency (FSA) \citep{fsa2007} and the World Health Organization (WHO) \citep{who2003diet}. Both provide tables based on a 2000kcal diet that contain ranges of nutrients, such as for example Fat, Saturated Fat, Sugar and Sodium (see Table \ref{tab:fsa} and Table \ref{tab:who}). The WHO guidelines account for macronutrients, such Fiber content, and so on. The FSA guidelines are typically used to derive front of package labels for meals and other food products sold in UK. In addition to the the nutrients per portion or per 100g, a traffic light system (red, amber, green) is used to inform the consumer, whether the meal is healthy (green) or unhealthy (red) with respect to a given property. We employed these guidelines in Table \ref{tab:rec_results2}. 
As the FSA scoring system is rather unpractical to use in a recommender scenario, one might want to use a single metric by following the procedure proposed by \citet{sacks2009impact} who first assign an integer value to each color (green=1, amber=2 and red=3) then sum the scores for each macro nutrient, resulting in a final range from 4 (very healthy) to 12 (very unhealthy). A further health index, which may offer utility is the `Healthy Eating Index' \citep{hei2016} proposed by the USDA. The index was developed to target the US population. To date it has not been applied in any food recommender systems project.

Other useful resources for building food recommender systems are provided by \textsc{foodsubs}\footnote{\url{http://www.foodsubs.com}}, a food thesaurus service which can suggest food substitutes. This might be helpful to implement food recommender systems promoting healthier eating (see \citep{AchananuparpW16}) by replacing unhealthy ingredients in a meal with more healthy variants, but also assist people with allergies or intolerances.

Food word lists, such as provided by \textsc{enchantedlearning}\footnote{\url{http://www.enchantedlearning.com/wordlist/food.shtml}} and \textsc{Wikipedia}\footnote{\url{http://www.wikipedia.org}} provides a rich knowledge base relating to food and cooking and may be used to assist with the normalization process of ingredients.

Finally, one may also employ health data as provided by the Centers for Disease Control and Prevention (CDC) in the US. The reports contain state and county data of diabetes and obesity. As different regions have different impact on what and how people eat (see \citep{trattner2017monitoring}), this might be a useful source of information when implementing food recommender systems for different regions and areas in the US \citep{said2014you}.

\subsection{Food Recommender System Frameworks}

To date, research in the food recommender systems domain relies mostly relies on software custom built by researchers themselves explicitly for the purpose of their research. To the best of our knowledge, there is no food recommender systems framework available that has been shared by the research community or on open-access platforms, such as \textsc{Github}\footnote{\url{http://www.github.com}}. This makes it challenging not only to progress the research in that area, but also to reproduce or validate findings published already. To counter this trend, in our own research, we have recently started to use publicly available frameworks, such as the well-known LibRec library. The framework is implemented in the Java programming language and comprises a relatively complete set of standard recommender systems algorithms, such as UserKNN, ItemKNN, BPR, SVD++, and so on, to tackle the rating prediction and item ranking problem. 
In \citep{trattner2017investigating} we adopted the framework with pre- and post-filtering functions (as described in the previous Section) to re-rank items (in our case) recipes in terms of their healthiness. We are happy to share this code upon request. The framework can also be easily extended to the problem of recommending, e.g., recipes to a group of people as well as generating personalized meal plans. Other examples of frameworks in other programming languages may be found on Graham Jenson's Github page\footnote{\url{https://github.com/grahamjenson/list_of_recommender_systems}} as well as on the RecSys Wiki\footnote{\url{http://www.recsyswiki.com/wiki/Recommendation_Software}}.

\section{Historical Evolution and Versions of the System}

The earliest examples of food recommender systems were proposed by the case-based reasoning (CBR) community \citep{hammond1986chef, hinrichs1989strategies}. In contrast to current state-of-the-art food recommender approaches both employed planning algorithms taking a set of queries e.g. groceries as input to generate meal plans or a single new recipe. Technically speaking these systems bear little relation to modern systems. Later, systems emerged employing simple variants of today's well-known content-based and collaborative filtering recommender algorithms. Examples include, for instance the works of \citet{lawrence2001personalization,mankoff2002using,aberg2006dealing}.

The first food recommenders built which are directly comparable to modern systems, i.e. which employ  standard algorithms such as UserKNN was presented in \citet{Freyne:2010:IFP:1719970.1720021, berkovsky2010group}. These were the first examples, where recipe datasets were used as a basis and the system was reliably evaluated. Subsequently other works emerged employing more advanced techniques to recommend food to people. Examples include the work of \citep{vanPinxteren:2011:DRS:1943403.1943422}, which was the first to derive a similarity metric for recipes to be used for recommending healthful meals;   \citep{Ueta:2011:RRS:2186633.2186642} and \citep{el2012food}, which employ knowledge-base food recommendation approaches; and  \citep{kuo2012intelligent} which employs tags to derive a knowledge graph to connect recipes and exploit this graph for recommending menus.

Other break through work was performed by
\citep{teng2012recipe}, who proposed the use of ingredient networks to produce recommendations or the work of \citet{Harvey:2013:YYE:2651320.2651339}, who proposed a model accounting for food selection biases.

A significant break-through was recently made by \citet{yang2017yum} who were able to develop a constraint-based (with different types of diets) mobile food recommender system exploring food images to learn about user food preferences. All previous approaches had relied on ratings or to some extent on tags \citep{Ge:2015:UTL:2750511.2750528}.

Behavior-based investigations, which go beyond the classic food recommender systems papers can also be considered to have progressed the field. We include our own work showing that people typically prefer the unhealthy recipes in this bracket \citep{trattner2017investigating}. This was the first study in the context that deals with the health-aware recipe recommender systems problem. Other work in this direction include \citep{Trattner:2017:RCI:3099023.3099072} (not shown in Table \ref{tbl:overview}) and \citep{Rokicki:2016:PPG:2930238.2930248} which illustrate differences in online food consumption with respect to hobbies and gender.

Finally, we would like to highlight our most recent work \citep{Elsweiler:2017:EFC:3077136.3080826} which investigated to which extent food recommender can nudge people towards healthier food choices.

\section{Evaluation: Metrics and Methodologies}\label{sec:eval}

The methods of evaluation applied to food recommender systems have evolved over time. 
The early concept papers found in the literature do not employ any kind of evaluation \citep{hammond1986chef,hinrichs1991roles}.
With the work of \citep{Freyne:2010:IFP:1719970.1720021} researchers started to employ evaluation techniques recognized by the community today as standard practices \citep{herlocker2004evaluating, ricci2011introduction}.

The most commonly taken approach (as can be seen in the summarized literature in Table \ref{tbl:overview}) is to perform simulations using historical data (see Section `Implementation Resources'). The experimental design specifics vary, but typically datasets are split into training and testing subsets to mimic user-profiles and feedback given for recommendations. Similar to other recommender domains, historical datasets are typically split such that 80\% of the data is used for training with the remaining 20\% held-out for testing. Alternatives are to use k-fold validation \citep{Harvey:2013:YYE:2651320.2651339,trattner2017investigating} or leave-one out protocol \citep{Freyne:2010:IFP:1719970.1720021}. The exact means by which collections are sourced varies from using naturalistic collections crawled from the web \citep{trattner2017investigating} or from donated sources \citep{Trevisiol:2014:BAR:2631775.2631784} to running user studies to collect small sets of data \citep{Harvey:2013:YYE:2651320.2651339}.

Different metrics have been applied to measure the performance of algorithms in such systems. These typically reflect the error in the predicted ratings \citep{Freyne:2010:IFP:1719970.1720021,Harvey:2013:YYE:2651320.2651339} e.g. Mean Absolute Error (MAE) or Root Spare Mean Error (RSME) or the quality of the top-n ranked list of items e.g. Recall, Precision,  Mean Average Precision (MAP) and Normalized Discounted Cumulative Gain (NDCG) \citep{Herlocker:2004:ECF:963770.963772}.

Mirroring the developments in the recommender systems community generally, earlier contributions focused on the rating prediction task whereas more recent and current work treats recommendation as a ranking problem (e.g., \citep{trattner2017investigating, yang2017yum, Cheng:2017:ICS:3079628.3079641}).

Assessing the accuracy of recommendation is typically not enough for recommender systems and in food recommenders is no exception.  Diversity of ingredients used in profiles was measured using Simpson and simple diversity metrics \citep{Elsweiler:2015:TAM:2792838.2799665}.

Incorporating health-aspects in the process requires additional metrics to be defined. As our own work shows, see \citep{trattner2017investigating}, metrics derived from the guidelines published by governmental bodies or health organizations are appropriate. 


In addition to calculating a mean over all food items recommended on per user basis (see Table \ref{tab:rec_results2}), we additionally introduced two further measures referred to as $\Delta$FSA and $\Delta$WHO, which capture the difference in healthfulness between test set items of a user and actual predicted items, as shown in the  formulae below
\begin{equation}
\Delta WHO = \sum_{u=1}^{|U|}\left(\sum_{i=1}^{|Train_u|} who_i -	\sum_{j=1}^{|Pred_u|} who_j\right)
\end{equation}
\begin{equation}
\Delta FSA = \sum_{u=1}^{|U|}\left(\sum_{i=1}^{|Train_u|} fsa_i -	\sum_{j=1}^{|Pred_u|} fsa_j\right)
\end{equation}
, where $|U|$ denotes the total number of users in the dataset, $|Train_u|$ the size of the train set for user $u$ respectively, $|Pred_u|$ the size of the set for the predicted items and $who_i$, $who_j$  and $fsa_i$, $fsa_j$ represent the WHO, FSA health scores for items ($i$ and $j$) in these sets. 

These delta measures are useful as they capture whether the recommended items are more or less healthy than those already rated positively by the user. The same procedure can also be applied to calculate a delta between the test and prediction sets to observe whether the recommended items are actually more or less healthy to what the user would actually eat in the future.

Similar to other recommender domains, studies employing online evaluation protocols, such as A/B testing or laboratory studies for the purpose  of testing the performance of food recommender systems are rare. Among the studies to employ online testing is for instance the work of \citep{freyne2011personalized} who ran two types of meal planners in a live system. The two methods tested were a personalized and a non-personalized algorithm. Over the course of 12 weeks over 5000 users participated in the study. According to the authors an A/B like setup was chosen to refer half of the users to the personalized condition and half of the user to the non-personalized one. 
Earlier work from the same authors employed also some variant of online experiment to gather ratings from users on recipes by e.g. using Amazon's Mechanical Turk platform \citep{freyne2011recipe,Freyne:2010:IFP:1719970.1720021}. However, rather than generating the recommendations on the fly to test their validity, in the end, an offline protocol was utilized. A recent study by \citet{yang2017yum} employed not only  offline testing but also an online study protocol to evaluate a mobile food recommender system. In particular they recruited 60 participants through the university mailing list, Facebook, and Twitter.  The study, conducted as a online Web service, consisted of three phases.  First, each participant was questioned on any dietary restrictions that may apply, such as the need to avoid gluten. Second, each user was asked to express their preferences by highlighting images of food they find appealing. Lastly, 20 meal recommendations were generated of which 10 were shown in a random order and 10 as proposed by the the authors' ``Yum-me'' algorithm. The participants had the task of classifying the 20 recipes as to whether it is appealing or not.

A final work worthy of mention is an online study that has been recently conducted by the authors with the goal of investigating the potential to nudge people towards healthier food choices via recommendations \citep{Elsweiler:2017:EFC:3077136.3080826}. The work employed three online studies. Similar to the previously mentioned work we implemented a Web service and recruited between 107 and 138 participants per study. By varying the amount of information shown about two algorithmically determined similar recipes, we were able to learn about the choices people make, the users' perception of these recipes and what influenced these. By applying machine learning approaches we were able to predict with relative certainty, which recipe of the two participants would prefer and demonstrate that the models developed can be used to influence the choices made.

In summary, no specialized offline protocols exist for the evaluation food recommender systems. Typically standard metrics are used to determine prediction accuracy and diversity. Furthermore, no standardized or specialized online evaluation protocols exist for food recommender systems. Current approaches rely on methods that have been previously developed in other recommender domains such as movies or music. Exceptions are the metrics specifically designed to incorporate healthy nutrition into the process, such as the WHO and FSA scores in \citep{trattner2017investigating}.

\section{Lessons Learned and Future Directions}\label{sec:lessons}

Thus, food recommendation is an important domain both for individuals and society. What the work described in this paper shows is that despite its importance, food item recommendation, in comparison to other domains is relatively under-researched. The work that has been performed to date shows that although user taste predictions for food can be achieved with existing methods, the performance achieved is poorer than in other domains.

This means that preference learning is should remain a focus for the food domain because experiments described in the literature have shown that even regardless of the source of user feedback applied (i.e. ratings, tags or comments) standard methods are only capable of producing relatively unsatisfactory performance. It is clear that new methods are required for the food domain and some work has shown promise. Yang and colleague's (2017) work uses images and embeddings (DNNs) to learn user preferences and the results are very promising.

Other key findings in the literature relating to preference prediction are those illustrating the importance of context variables. One promising research direction would be to capture important context variables via different sensors and incorporate these into recommendation models.

Relating to context, social situations and recommendation for groups needs to be considered more concretely. The pervasiveness of social culinary experiences and how these influence food choices need to be considered by technological systems.

One particular task in food recommender systems, which for societal and socio-economic reasons, has become a hot research focus is food recommenders for nutritional health. Researchers have proposed diverse methods of incorporating  nutrition (nutritional components in algorithm, meal plans, and nudging), but to date all of these proposals remain preliminary and it is not yet clear, which is the best approach to take.

As a final note, one further aspect which needs to develop in the community is the evaluation of food recommenders and the methods employed to do so. In the literature evaluation has mostly been offline with proprietary collections. As a community we need to work together to achieve standard data collections, standard base-line approaches and importantly, more online studies to understand how our approaches work as live systems used in naturalistic scenarios.

\bibliographystyle{apacite}    
\bibliography{references}      

\end{document}